\documentclass[dvips]{PoS}
\usepackage{amsmath,amssymb}
\usepackage{graphicx}
\usepackage{subfigure}

\newcommand{\mps}{m_\mathrm{PS}}
\newcommand{\fps}{f_\mathrm{PS}}

\newcommand{\mev}{\mathrm{MeV}}
\newcommand{\preprint}{\newline%
  \begin{picture}(0,0)
  \put(300,112){\rm\small MS-TP-10-16, SFB/CPP-10-97}
  \end{picture}}

\title{Pseudoscalar decay constants from $N_f=2+1+1$ twisted mass
  lattice QCD\preprint}

\ShortTitle{Pseudoscalar decay constants from $N_f=2+1+1$ tmLQCD}

\author{Federico Farchioni\\
  Universit\"at M\"unster, Institut f\"ur Theoretische Physik,\\
  Wilhelm-Klemm-Stra{\ss}e 9, 48149 M\"unster, Germany\\
  E-mail: \email{farchion@uni-muenster.de}
}
\author{Gregorio Herdoiza, Karl Jansen, Andreas Nube\\
  DESY, Platanenallee 6, 15738 Zeuthen, Germany\\
  E-mail: \email{gregorio.herdoiza@desy.de}, \email{karl.jansen@desy.de}, \email{andreas.nube@desy.de}\\
}
\author{Marcus Petschlies\\
  Humboldt Universit{\"a}t zu Berlin, Institut f{\"u}r Physik\\
  Newtonstra{\ss}e 15, 12489 Berlin, Germany\\
  E-mail: \email{marcus.petschlies@physik.hu-berlin.de}
}
\author{\speaker{Carsten Urbach}\\
  Helmholtz Institut f{\"u}r Strahlen und Kernphysik and
  Bethe Center for Theoretical Physics,\\
  Universit{\"a}t Bonn, Nussallee 14-16, 53115 Bonn, Germany\\
  E-mail: \email{urbach@hiskp.uni-bonn.de}
}

\abstract{We present first results for the pseudoscalar decay
  constants $f_K$, $f_D$ and $f_{D_s}$ from lattice QCD with
  $N_f=2+1+1$ flavours of dynamical quarks. The lattice simulations
  have been performed by the European Twisted Mass collaboration
  (ETMC) using maximally twisted mass quarks. For the pseudoscalar
  decay constants we follow a mixed action approach by using so called
  Osterwalder-Seiler fermions in the valence sector for strange and
  charm quarks. The data for two values of the lattice spacing and
  several values of the up/down quark mass is analysed using chiral
  perturbation theory.}

\FullConference{The XXVIII International Symposium on Lattice Field Theory, Lattice2010\\
		June 14-19, 2010\\
		Villasimius, Italy}

\begin{document}

\section{Introduction}

The decay constants of the Pion, Kaon, the $D$- and the $D_s$-meson
are phenomenologically interesting quantities, not least because the
ratio $f_K/f_\pi$ together with the well known $|V_{ud}|$ can be used
to determine $|V_{us}|$. While experimentally $f_\pi$ and $f_K$ are
well known and $f_D$ and $f_{D_s}$ less so, lattice QCD is in
principle able to provide calculations of all of these from first
principles. And recent advances in the field allow now also for
statistically precise determinations with $N_f=2$ and $N_f=2+1$
dynamical quark
flavours~\cite{Aubin:2004fs,Follana:2007uv,Allton:2008pn,Blossier:2009bx,Durr:2010hr},
with systematic uncertainties more or less under control.

In this proceeding contribution we are going to present yet another
determination of the aforementioned decay constants, however with a
dynamical charm quark in place, i.e. for QCD with $N_f=2+1+1$ quark
flavours. 

\begin{table}[t!]
  \centering
  \begin{tabular*}{.8\textwidth}{@{\extracolsep{\fill}}lccccc}
    \hline\hline
    ensemble & $\beta$ & $a\mu_\ell$ & $a\mu_\sigma$ & $a\mu_\delta$ &
    $L/a$ \\
    \hline\hline
    B35.32 & $1.95$ & $0.0035$ & $0.135$ & $0.170$ & $32$ \\ 
    B55.32 & $1.95$ & $0.0055$ & $0.135$ & $0.170$ & $32$ \\ 
    B75.32 & $1.95$ & $0.0075$ & $0.135$ & $0.170$ & $32$ \\ 
    B85.24 & $1.95$ & $0.0085$ & $0.135$ & $0.170$ & $24$ \\
    \hline
    D20.48 & $2.10$ & $0.0020$ & $0.120$ & $0.1385$ & $48$ \\
    D30.48 & $2.10$ & $0.0030$ & $0.120$ & $0.1385$ & $48$ \\
    \hline\hline
  \end{tabular*}
  \caption{The ensembles used in this investigation. The notation of
    ref.~\cite{Baron:2010bv} is used for labeling the ensembles.}
  \label{tab:setup}
\end{table}

\section{Set-up}

We use gauge configurations as produced by the European Twisted Mass
Collaboration (ETMC) with $N_f=2+1+1$ flavours of Wilson twisted mass
quarks and Iwasaki gauge action. The set-up is described in
ref.~\cite{Baron:2010bv} and the ensembles used in this investigation
are summarised in table~\ref{tab:setup}. The twisted mass Dirac
operator in the light -- i.e. up/down -- sector
reads~\cite{Frezzotti:2000nk}
\begin{equation}
  \label{eq:Dud}
  D_\ell = D_W + m_0 + i \mu_\ell \gamma_5\tau_3
\end{equation}
and in the strange/charm sector~\cite{Frezzotti:2003xj}
\begin{equation}
  \label{eq:Dsc}
  D_h = D_W + m_0 + i \mu_\sigma \gamma_5\tau_1 + \mu_\delta \tau_3\, ,
\end{equation}
where $D_W$ is the Wilson Dirac operator. The value of $m_0$ was tuned
to its critical value as discussed in
refs.~\cite{Chiarappa:2006ae,Baron:2010bv} in 
order to realise automatic $\mathcal{O}(a)$ improvement at maximal
twist~\cite{Frezzotti:2003ni}. Note that the bare twisted masses
$\mu_{\sigma,\delta}$ are related to the bare strange and charm quark
masses via the relation
\begin{equation}
  \label{eq:msc}
  m_{c,s} = \mu_\sigma\ \pm\ (Z_\mathrm{P}/Z_\mathrm{S})\ \mu_\delta
\end{equation}
with pseudoscalar and scalar renormalisation constants $Z_\mathrm{P}$
and $Z_\mathrm{S}$.

In the strange and charm quark sector, we use a mixed action approach with
Osterwalder-Seiler (OS) valence quarks by formally introducing a twisted
doublet for valence strange and charm quark
flavours~\cite{Frezzotti:2004wz,Blossier:2007vv}, a set-up without
flavour mixing artifacts in the valence sector. The two actions in
the sea and the valence sector can be matched by tuning the bare values
of valence strange and charm such that unitary Kaon and D-meson masses
are reproduced. Unfortunately, it turned out that the unitary Kaon and
D-meson masses were not exactly tuned to their physical values and
currently we are still lacking ensembles which would allow to
interpolate to the corresponding physical values. Hence, for the time
being we vary the valence quark masses to interpolate the valence
Kaon and D-meson masses to their physical values, as discussed below. 
This approach has been successfully applied to the ETMC $N_f=2$
flavour gauge configurations in ref.~\cite{Blossier:2009bx}.

The determination of the (unitary) K-meson mass is described in
ref.~\cite{Baron:2010th}. Its decay constant can be determined from
\begin{equation}
  \label{eq:fKunitary}
  f_K = (m_\ell + m_s) \frac{ \langle 0 | \tilde P_K | K\rangle}{m_K^2}\, ,
\end{equation}
where $\tilde P_K$ represents the projection to the physical Kaon
interpolating operator as discussed in
ref~\cite{Baron:2010th}. $m_{s}$ is the strange quark masses defined in
eq.~(\ref{eq:msc}). The pseudoscalar decay constant $\fps$ in the
valence sector is determined from 
\begin{equation}
  \label{eq:fPS}
  \fps = \left(\mu_{val}^{(1)} + \mu_{val}^{(2)}\right) \, \frac{\vert \langle 0
    \vert P \vert PS \rangle\vert}{\mps\,\sinh \mps} \, ,  
\end{equation}
where $P=\bar q_1 \gamma_5 q_2$ with quark fields $q_1,q_2$ suitably
chosen for the desired quark content. The meson mass $\mps$ and the matrix element
$\vert\langle 0 \vert P \vert PS \rangle\vert$ entering eq.~(\ref{eq:fPS}) have
been extracted from a single state fit of the corresponding two-point
pseudoscalar correlation function. The replacement of $\mps$ with
$\sinh \mps$ in the lattice definition~(\ref{eq:fPS}) of the decay
constant helps in reducing discretisation errors for heavy meson
masses~\cite{Blossier:2009bx}. 

\begin{table}[t!]
  \centering
  \begin{tabular*}{.8\textwidth}{@{\extracolsep{\fill}}lcc}
    \hline\hline
    & $\beta=1.95$ & $\beta=2.10$\\
    \hline\hline
    $a\mu_s$ & $0.0130$ & $0.0110$ \\
             & $0.0145$ & $0.0120$ \\
             & $0.0160$ & $0.0130$ \\
             & $0.0180$ & $0.0150$ \\
             & $0.0210$ & $0.0180$ \\
             &          & $0.0200$ \\
    \hline\hline
  \end{tabular*}
  \caption{Bare values of the valence quark masses in the strange
    region for $\beta=1.95$ and $\beta=2.10$.}
  \label{tab:mus}
\end{table}

Any value for $\fps$ and $\mps$ depends on the mass-value of the light
dynamical quark (sea strange and charm quark mass values are fixed for a
given $\beta$-value) and two valence quark masses, which we denote by
$\fps(\mu_\ell^{sea}, \mu^{val}_1, \mu^{val}_2)$. We have investigated
several values for the valence quarks: $\mu^{val} = \mu^{sea}_\ell =
\mu_\ell$ and five to six values in the strange and charm quark region
$\mu^{val} = \mu_{s,c}$. The ensembles used are summarised in
table~\ref{tab:setup} and the bare values of the valence strange and
charm quark masses in tables~\ref{tab:mus} and \ref{tab:muc}.

\begin{table}[t!]
  \centering
  \begin{tabular*}{.8\textwidth}{@{\extracolsep{\fill}}lcc}
    \hline\hline
    & $\beta=1.95$ & $\beta=2.10$\\
    \hline\hline
    $a\mu_c$ & $0.200$ & $0.1650$ \\
             & $0.215$ & $0.1800$ \\
             & $0.240$ & $0.2000$ \\
             & $0.260$ & $0.2250$ \\
             &         & $0.2500$ \\
    \hline\hline
  \end{tabular*}
  \caption{Bare values of the valence quark masses in the charm region
    for $\beta=1.95$ and $\beta=2.10$.}
  \label{tab:muc}
\end{table}

\section{Results}

\begin{figure}[t]
  \centering
  \includegraphics[width=0.7\linewidth]{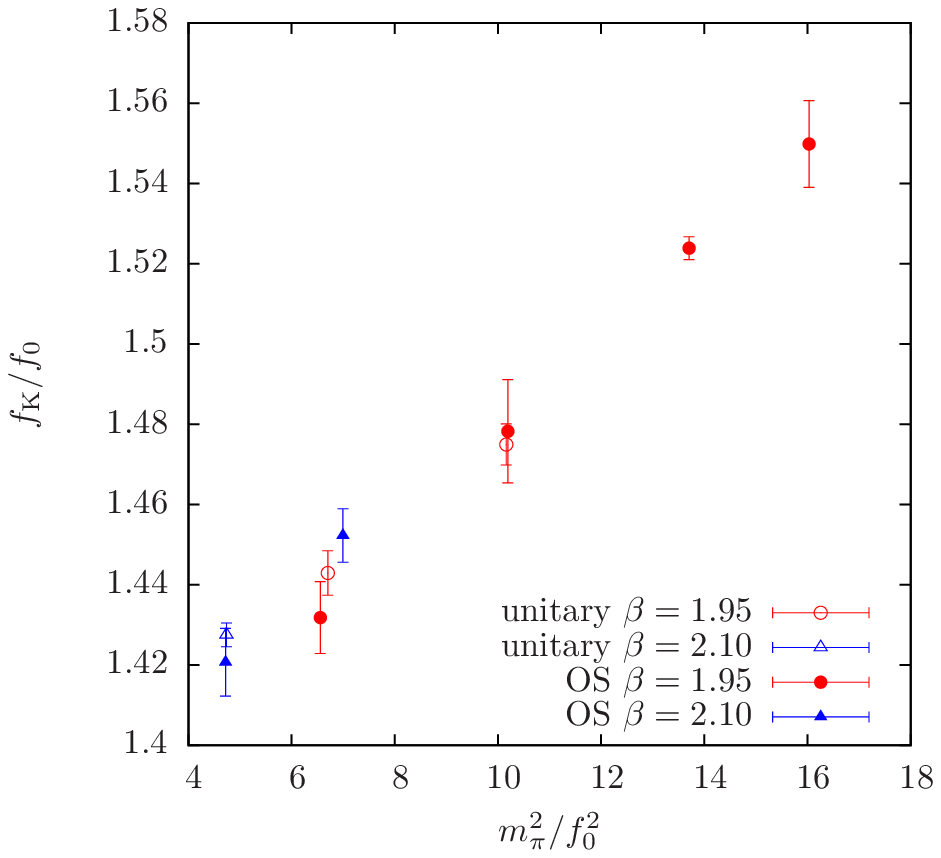}
  \caption{$f_K/f_0$ as a function of the pion mass squared. Results
    for the unitary and the mixed action $f_K$ are shown for two
    values of the lattice spacing.}
  \label{fig:compfK}
\end{figure}

One may expect large differences in between the unitary set-up and the
mixed action set-up at finite values of the lattice spacing. In order
to investigate this point we have determined $f_K$ in the unitary and
the mixed action set-up after matching the Kaon mass. The result is
shown in figure~\ref{fig:compfK} where we plot $f_K$ as a function of
the squared pion mass. Both quantities are in units of $f_0$, the pion
decay constant in the chiral limit as determined in
ref.~\cite{Baron:2010bv}. Within errors unitary and mixed set-up
determination of $f_K$ agree. Moreover, results for the two available
values of the lattice spacing agree within errors. This indicates
small lattice and small unitarity breaking artifacts at least in
$f_K$. Note that the ratio $Z_\mathrm{P}/Z_\mathrm{S}$ used for the
determination of the unitary $f_K$ has been determined by matching the
bare value of the OS strange quark mass to $\mu_\sigma$ and
$\mu_\delta$ via eq.~(\ref{eq:msc}).

In order to determine the physical value of $f_K$ we fit SU(2)
$\chi$PT formulae to our pion and Kaon decay constants
data~\cite{Gasser:1983yg,Allton:2008pn} simultaneously according to 
\begin{eqnarray}
  \label{eq:fpifit}
  \fps(\mu_\ell,\mu_\ell,\mu_\ell) &=& f_0 \cdot \left( 1 - 2\,\xi_{ll}\ln \xi_{ll} +
    b\, \xi_{ll} \right)\,, \\
  \fps(\mu_\ell,\mu_\ell,\mu_s) &=& (f^{(K)}_0 + f^{(K)}_m\, \xi_{ss}) \cdot \nonumber \\
  && \label{eq:fkfit}
  \cdot \left[ 1 - \dfrac{3}{4} \xi_{ll} \ln\xi_{ll} + (b^{(K)}_0 + b^{(K)}_m\,
    \xi_{ss})\,\xi_{ll}
  \right] \,
\end{eqnarray}
where
\begin{equation}
  \label{eq:xi}
  \xi_{XY} = \frac{m_{PS}^2(\mu_\ell,\mu_X,\mu_Y)}{(4\pi f_0)^2}\,
\end{equation}
are expressed in our analysis as a
function of meson masses\footnote{We use the normalization in which
$f_\pi=130.7\, \mev$.}. We correct our data for finite size effects
using NLO $\chi$PT~\cite{Gasser:1987ah,Becirevic:2003wk}
\begin{equation}
  \label{eq:ffse}
  \begin{split}
    \fps(\mu_\ell,\mu_\ell,\mu_\ell;L) &= \fps(\mu_\ell,\mu_\ell,\mu_\ell)  \cdot \left[ 1 -
      2\,\xi_{ll}\, \tilde g_1(L, \xi_{ll}) \right] \,,\\
    \fps(\mu_\ell,\mu_\ell,\mu_s;L) &= \fps(\mu_\ell,\mu_\ell,\mu_s)  \cdot \left[ 1 -
      \dfrac{3}{4}\,\xi_{ll}\, \tilde g_1(L, \xi_{ll}) \right] \, .\\
  \end{split}
\end{equation}
We fit to our $\beta=1.95$ and $\beta=2.10$ data simultaneously. We do
not include lattice artifacts of 
$\mathcal{O}(a^2)$ in our fit, because the amount of data is not
sufficient to do so: including these effects lets the fits become
instable. Moreover, the data is fittable without these terms with
$\chi^2/\mathrm{dof} = 50/30$. 

\begin{figure}[t]
  \centering
  \includegraphics[width=0.7\linewidth]{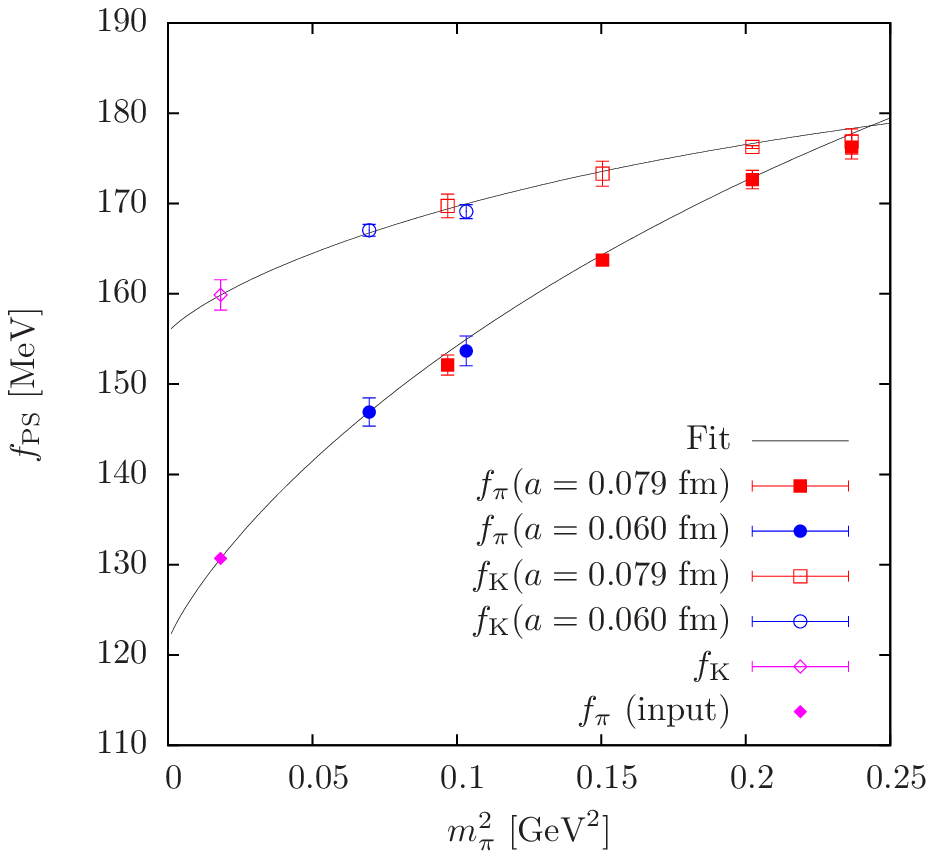}
  \caption{Pion and Kaon decay constants as a function of
    $m_\pi^2$. We show data for two values of the lattice spacing
    $a=0.079\ \mathrm{fm}$ and $a=0.060\ \mathrm{fm}$, corresponding
    to $\beta=1.95$ and $\beta=2.10$.}
  \label{fig:fKfpi}
\end{figure}

The physical input to our fits is $f_\pi=130.7\ \mev$, $m_\pi=135\ \mev$ and
$m_K=497.7\ \mev$. In figure~\ref{fig:fKfpi} we show pion and Kaon
decay constants as a function of the squared pion mass $m_\pi^2$. The
corresponding Kaon decay constant is determined at a value of
$\mps^2(\mu_\ell, \mu_s, \mu_s)
= 2m_K^2 - m_\pi^2$. For the figure we have interpolated our data
linearly to these values. As a result we obtain $f_K/f_\pi =
1.224(13)$, $f_K = 160(2)\ \mathrm{MeV}$ and $\bar\ell_4 = 4.78(2)$
with statistical errors only as determined from a bootstrap analysis. 

The fit also allows us to determine the values of the lattice spacings
at $\beta=1.95$ and $\beta=2.10$ and the value of $f_0$. All these
quantities agree very well with the results obtained in
ref.~\cite{Baron:2010bv}. 

Following the procedure described in ref.~\cite{Blossier:2009bx} we
have also analysed the data for $f_D$ and $f_{D_s}$ using SU$(2)$
heavy meson chiral perturbation theory~\cite{Sharpe:1995qp}. We
consider here the expansions of 
\begin{equation}
  \label{eq:fD}
  f_{D_s}\sqrt{m_{D_s}}\quad \textrm{and}
  \quad\frac{f_{D_s}\sqrt{m_{D_s}}}{f_{D}\sqrt{m_{D}}} 
\end{equation}
including terms proportional $a^2 m_{D_s}^2$ and $1/m_{D_s}$. For
details we refer to ref.~\cite{Blossier:2009bx}. Our first results for
$f_D$ and $f_{D_s}$ are very encouraging, however, quoting results
requires a better control of the systematics involved in this
investigation.

\section{Summary and Outlook}

We have presented the first determination of the pseudoscalar decay
constants $f_K$, $f_D$ and $f_{D_s}$ from lattice QCD with dynamical
up, down, strange and charm quark flavours. We have used a mixed
action approach with Osterwalder-Seiler valence quarks on a maximally
twisted mass sea. The analysis indicates small lattice and
unitarity breaking artifacts. The preliminary results are $f_K/f_\pi =
1.224(13)$ and $f_K = 160(2)\ \mathrm{MeV}$ obtained with a SU$(2)$
chiral perturbation theory fit. The errors are statistical only. Using
the results of ref.~\cite{Marciano:2004uf} this translates to
$|V_{us}|=0.220(2)$. 

A comparison of our results for $f_K$ and $f_K/f_\pi$ to results
available in the literature shows that to the current level of
accuracy there is no difference visible, neither to $N_f=2$ flavour
results~\cite{Blossier:2009bx} nor to $N_f=2+1$ flavour
results~\cite{Allton:2008pn,Follana:2007uv,Aubin:2004fs}. 

A similar analysis has been performed for $f_D$ and $f_{D_s}$. In the
case of these charmed quantities it will be in particular interesting
to understand whether lattice artifacts proportional to $a^2m_c^2$ are
small enough in our lattice set-up to allow for a precise
determination of the corresponding quantities. However, the $N_f=2$
results presented in ref.~\cite{Blossier:2009bx} give rise to optimism
that also $f_D$ and $f_{D_s}$ can be reliably determined in our set-up. 

Clearly the results presented here need a better understanding of the
systematic uncertainties. This includes in particular the dependence
of the decay constants on the sea strange and charm quark mass
values. ETMC is currently producing ensembles to investigate this
point. These new ensembles should eventually allow to interpolate to
the physical values of $m_K$ and $m_D$. 
Moreover, more ensembles with different light quark mass values
are required for $\beta=2.10$ to better control lattice
artifacts, for which also a third value of the lattice spacing is
desirable. 

On the analysis site we are currently implementing different fit
formulae (see for instance refs.~\cite{Durr:2010hr,Bernard:2010ex}) in
order to better understand the extrapolation in the various quark masses.

\subsection*{Acknowledgements}

We thank Marc Wagner for useful discussions. We thank the members of ETMC
for the most enjoyable collaboration. The computer time for this
project was made available to us by the John von Neumann-Institute for
Computing (NIC) on the JUMP, Juropa and Jugene systems in Jülich and
apeNEXT system in Zeuthen, BG/P and BG/L in Groningen, by BSC on
Mare-Nostrum in Barcelona (www.bsc.es) and by the computer resources
made available by CNRS on the BlueGene system at GENCI-IDRIS 
Grant 2009-052271 and CCIN2P3 in Lyon. We thank these computer centers
and their staff for all technical advice and help. This work has been
supported in part by the DFG Sonderforschungsbereich TR9
Computergestützte Theoretische Teilchenphysik and the EU Integrated
Infrastructure Initiative Hadron Physics (I3HP) under contract
RII3-CT-2004-506078.

\bibliographystyle{h-physrev5}
\bibliography{bibliography}

\end{document}